\documentclass[10pt,twocolumn,letterpaper]{article}
\usepackage{mathtools}
\usepackage{cvpr}              
\usepackage[dvipsnames]{xcolor}

\definecolor{cvprblue}{rgb}{0.21,0.49,0.74}
\usepackage[pagebackref,breaklinks,colorlinks,citecolor=cvprblue]{hyperref}
\usepackage{colortbl}
\definecolor{tabfirst}{rgb}{1, 0.7, 0.7} 
\definecolor{tabsecond}{rgb}{1, 0.85, 0.7} 
\definecolor{tabthird}{rgb}{1, 1, 0.7} 

\title{ConVRT: Consistent Video Restoration Through Turbulence with \\ Test-time Optimization of Neural Video Representations}

\author{Haoming Cai$^{*1}$,\quad Jingxi Chen\thanks{Equal Contribution} $^{1}$,\quad Brandon Y. Feng$^{2}$,\quad Weiyun Jiang$^{3}$,\\ Mingyang Xie$^{1}$,\quad Kevin Zhang$^{1}$,\quad Ashok Veeraraghavan$^{3}$,\quad Christopher Metzler\thanks{Corresponding author.} $^{1}$\\
$^{1}$University of Maryland, College Park\quad 
$^{2}$Massachusetts Institute of Technology\quad 
$^{3}$Rice University
\and
\url{https://convrt-2024.github.io/}
}
\begin{document}
\maketitle

\begin{abstract}
Atmospheric turbulence presents a significant challenge in long-range imaging.
Current restoration algorithms often struggle with temporal inconsistency, as well as limited generalization ability across varying turbulence levels and scene content different than the training data.
To tackle these issues, we introduce a self-supervised method, \textbf{C}onsistent \textbf{V}ideo \textbf{R}estoration through \textbf{T}urbulence (ConVRT) a test-time optimization method featuring a neural video representation designed to enhance temporal consistency in restoration.
A key innovation of ConVRT is the integration of a pretrained vision-language model (CLIP) for semantic-oriented supervision, which steers the restoration towards sharp, photorealistic images in the CLIP latent space. 
We further develop a principled selection strategy of text prompts, based on their statistical correlation with a perceptual metric.
ConVRT's test-time optimization allows it to adapt to a wide range of real-world turbulence conditions, effectively leveraging the insights gained from pre-trained models on simulated data. ConVRT offers a comprehensive and effective solution for mitigating real-world turbulence in dynamic videos.
\end{abstract}

\section{Introduction}
\label{sec:intro}

Atmospheric turbulence often occurs in aerial photography and astronomical observations and significantly degrades imaging quality, leading to blurred, warped, or otherwise distorted imagery.
Effective turbulence mitigation is not only crucial for enhancing the clarity and reliability of visual information, but also plays a pivotal role in various applications ranging from remote sensing and security surveillance to scientific research and environmental monitoring.
\begin{figure}[t]
    \centering
    \includegraphics[width=\linewidth]{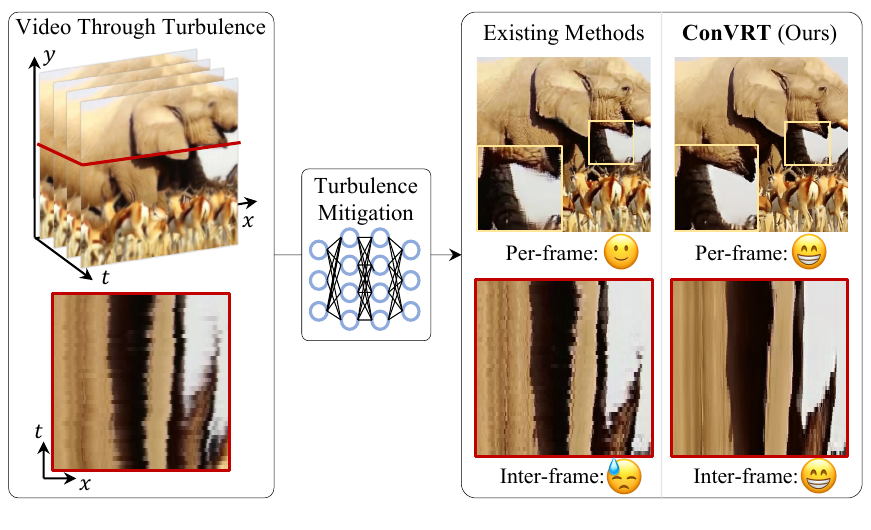}
    \caption{Existing turbulence mitigation methods produce good per-frame results but often fail to maintain consistency across frames, which is vital for downstream tasks. Our work introduces ConVRT, which effectively removes turbulence while preserving temporal consistency in the restored video.}
    \label{fig:intro}
\end{figure}

\begin{figure*}[!tbp]
    \centering
    \includegraphics[width=.95\linewidth]{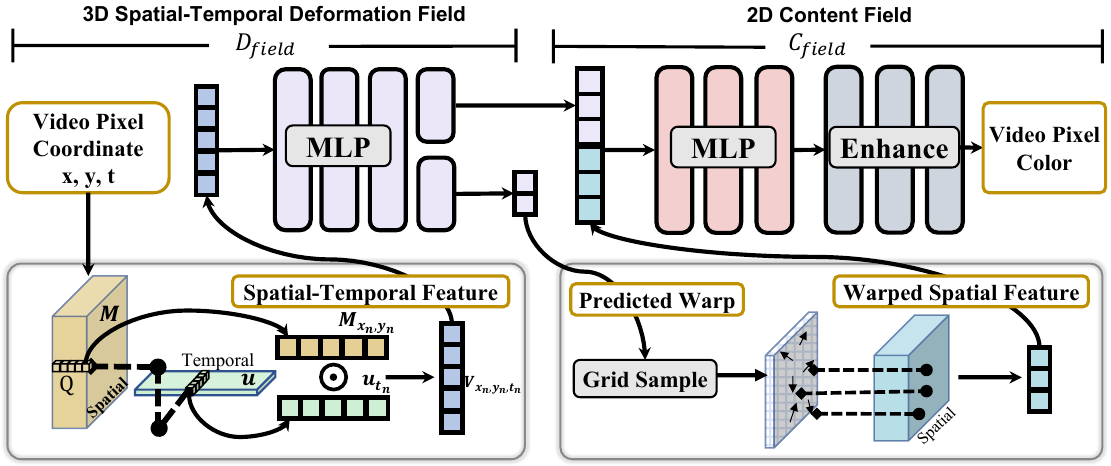}
    \caption{Illustration of the proposed method : ConVRT decomposes a video into two fields: 3D Spatial-Temporal Deformation Field \( D_{field} \) and 2D Content Field \( C_{field} \). Low-rank decomposition in \( D_{field} \) reduces parameters and preserves spatial-temporal details. The predicted warp from \( D_{field} \) shapes the spatial features in \( C_{field} \), affecting the RGB frame output.}
    \label{fig:sys_overview}
\end{figure*}

While turbulence mitigation on static scenes has seen remarkable advancements, largely fueled by the availability of extensive datasets, improved turbulence simulators, and the development of more capable machine learning algorithms, the domain of dynamic video restoration under turbulence conditions lags behind. This lag can be attributed to several unique challenges inherent to video processing.

A primary obstacle in video-based restoration is maintaining high temporal consistency.
Unlike static scenes, video observations of dynamic scenes comprise a sequence of frames where each frame is not only expected to be clear, but also consistent with other frames in terms of quality and continuity. 
This requirement for temporal coherence adds a layer of complexity to the restoration process.
Furthermore, turbulence distortions vary not only spatially across a single frame but also temporally across the sequence of frames, making the restoration process significantly more intricate.


The prevailing approaches to video turbulence mitigation typically involve either the iterative application of single-image restoration methods to each frame or the use of video-based restoration networks trained on simulated data. However, these strategies are not without significant drawbacks. When single-image methods are applied independently to each frame, they often fail to maintain inter-frame continuity, resulting in jittery or inconsistent visual outputs. Furthermore, the reliance on simulators for the development and testing of these methods introduces a notable accuracy gap. While simulators are useful for dataset generation, they may not accurately replicate the complex and dynamic nature of real-world atmospheric turbulence. As such, methods with good performance on simulated data do not always perform effectively in real-world scenarios.

These challenges highlight an urgent need for more robust, adaptable, and specialized approaches in video-based turbulence mitigation. A pivotal question arises: Is it possible to develop a method that not only leverages the valuable insights gained from pre-trained methods using simulated data, but also adapts to the constantly changing conditions of real-world turbulence during test time?

This paper presents, \textbf{C}onsistent \textbf{V}ideo \textbf{R}estoration through \textbf{T}urbulence (ConVRT), a strategy which combines the adaptability required for real-world application with the foundational strengths of simulation-based training.
Moving away from the conventional reliance on complex deep learning models or intricate turbulence simulators for inverse rendering, ConVRT innovatively combines recent advances in machine learning with neural signal representations to address video-based turbulence mitigation.

At the heart of ConVRT lies a neural video representation, which is composed of a 2D content field and a 3D spatial-temporal deformation field. 
This dual-field representation allows for a more nuanced and accurate restoration of video content distorted by atmospheric turbulence.
We employ a test-time optimization framework to train this video representation, effectively modeling both the dynamic scene and the turbulence distortions.
After optimization, the dynamic scene can decouple from the turbulence distortion, resulting in a sharply restored video.
To further refine the reconstruction, ConVRT incorporates semantic-oriented supervision using priors from the Contrastive Language-Image Pre-Training (CLIP) \cite{radford2021learning} model.
We propose a novel strategy that selects prompts based on the statistical correlation between CLIP and the LPIPS \cite{zhang2018unreasonable} metric, guiding the restoration process such that the output is more closely aligned with the ideal prompt within the projected embedding space of CLIP. 
Additionally, a key focus of ConVRT is on improving temporal consistency, achieved through the careful design of the neural representation of the deformation field. 
Crucially, our test-time optimization framework circumvents the typical generalization issues of deep-learning-based methods while retaining the flexibility to incorporate pretrained knowledge from existing models.

ConVRT addresses the key challenges of restoring video distorted by turbulence. The major contributions include:
\begin{itemize}[leftmargin=.3in]
\item An innovative test-time optimization framework for turbulence mitigation, improving per-frame restoration fidelity and inter-frame temporal coherence.

\item An efficient neural representation of videos tailored specifically for turbulence mitigation, including a pair of content field and deformation field.

\item A novel, semantic-oriented enhancement module using the pretrained CLIP model, including developing a principled strategy for prompt selection based on the statistical correlation between CLIP and LPIPS.

\item A comprehensive evaluation against existing methods, showing it outperforming existing methods on visual quality and coherence of the restored video content.

\end{itemize}

\section{Related Work}
\noindent \textbf{Implicit neural representations.}
Our work leverages a coordinate-based implicit neural representation (INRs), which has been commonly adopted to model 2D images or 3D videos as multi-layer perceptions (MLPs). INRs take 2D pixel coordinates $(x, y)$, or 3D pixel coordinates with temporal encoding, $(x, y, t)$ and output the corresponding pixel values. These INRs demonstrate exceptional performance when fitting images~\cite{mildenhall2021nerf, tancik2020fourier, mehta2021modulated, feng2021signet, feng2022viinter, feng2022neural, muller2022instant}, videos~\cite{sitzmann2020implicit, chen2021nerv, chen2023hnerv, attal2023hyperreel}, and 3D shapes~\cite{park2019deepsdf,feng2022prif, sitzmann2020implicit}. Not only they are able to represent these 2D or 3D signals, but they also show strong priors for solving inverse problems, such as image super resolution~\cite{chen2021learning}, video inpainting~\cite{chen2023hnerv}, phase retrieval~\cite{wang2023local, zhou2023fpm}, and reducing optical aberration~\cite{feng2023neuws, lin2023learning, bostan2020deep}.

\noindent \textbf{Neural video representation.}
Our work aligns closely with the evolving field of neural video representation~\cite{li2023dynibar, wang2023tracking, feng20233d, ouyang2023codef}. While there are existing approaches~\cite{kasten2021layered, lee2023shape, ouyang2023codef, ye2022deformable} that seek to represent a video into decomposed layers, these primarily focus on clean videos and are not applicable to videos with severe degradation turbulence. Our work extends the application of neural video representation to scenarios heavily impacted by atmospheric turbulence. This extension is not trivial, as it involves addressing the unique challenges posed by the dynamic and unpredictable nature of turbulence, which are not considered in conventional video representations.

\noindent \textbf{Atmospheric turbulence mitigation.}
Attempts to mitigate atmospheric turbulence~\cite{fried1978probability, noll1976zernike} have applied optical flow~\cite{mao2020image, caliskan2014atmospheric}, B-spline grid~\cite{shimizu2008super}, and diffeomorphism~\cite{gilles2008atmospheric} to unwarp each distorted image and then fuse and combine these registered distorted images into a clean and sharp image.
The fusion is usually modeled as patch-wise stitching~\cite{mao2020image} or blind deconvolution~\cite{anantrasirichai2018atmospheric}.
Recent development of high-performance GPUs and fast turbulence simulators leads to new progress in turbulence mitigation~\cite{chimitt2023anisoplanatic, chimitt2023scattering, mao2021accelerating, mao2022single, feng2022turbugan, zhang2022imaging, jiang2023nert}.
However, previous efforts tend to overlook the importance of temporal consistency on the reconstructed video. Our method, ConVRT, is specifically designed to restore temporal consistency with on test-time optimization of a neural video representation.

\begin{figure*}
  \centering
  \includegraphics[width=.96\linewidth]{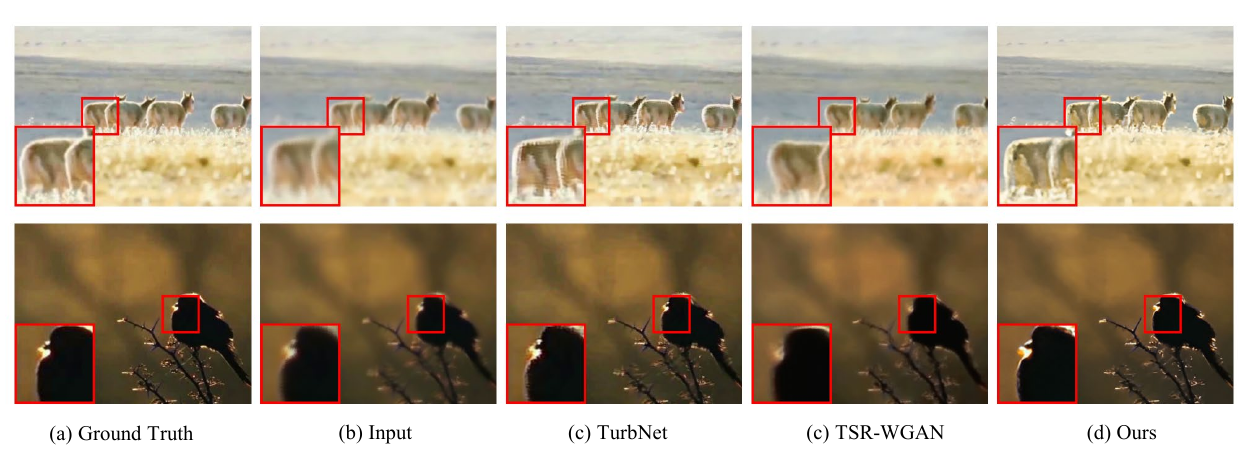}
    \caption{Comparing single-frame restoration in synthetic turbulence videos with TurbNet\cite{mao2022single} and TSR-WGAN\cite{jin2021neutralizing}: Our results are the sharpest and most accurate.}
    \label{fig:sim_single}
\end{figure*}


\begin{figure*}
  \centering
  \includegraphics[width=.96\linewidth]{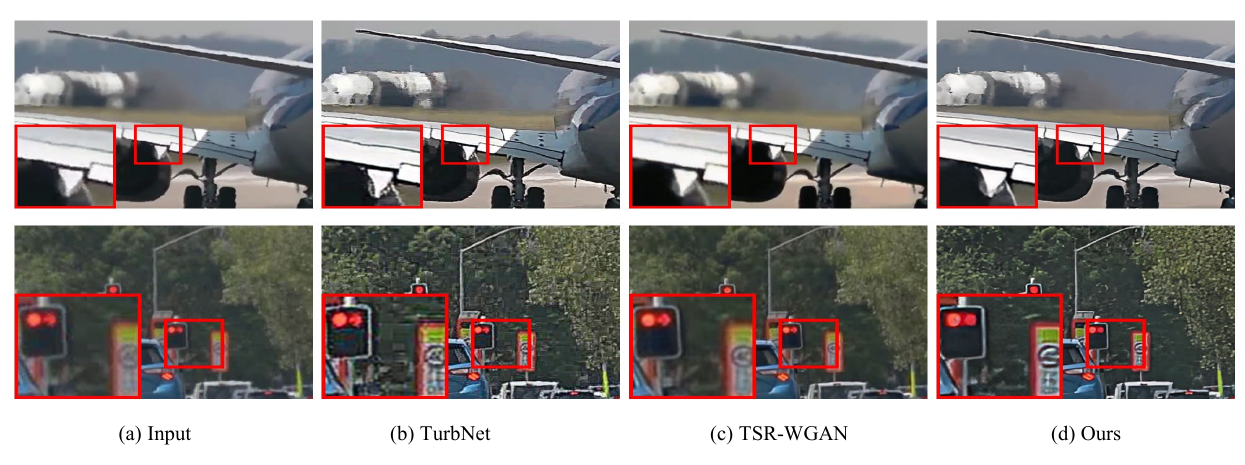}
    \caption{Comparison of single-frame restoration quality on real-world turbulence videos: TurbNet\cite{mao2022single} results are with distorted boundaries and artficts; TSR-WGAN\cite{jin2021neutralizing} results do not really remove the turbulence blur as shown on zoom-in views.}
    \label{fig:real_single}
\end{figure*}

\section{Method}
This section describes the proposed design of a neural video representation tailored to turbulence mitigation.

\subsection{General Pipeline}
\label{subsec:general_pipeline}
The framework of our method, ConVRT, is presented in Figure~\ref{fig:sys_overview}. 
During training, TurbNet's output \cite{mao2022single} serves as the sole supervision signal for our ConVRT. 
ConVRT is designed to adapt to future advancements in turbulence mitigation algorithms.
As for the pipeline design, ConVRT employs a dual-field approach: a 3D Spatial-Temporal Deformation Field \( D_{field} \) for adapting to temporal variations, and a 2D Content Field \( C_{field} \) for canonical 2D content. 
\( D_{field} \) generates spatial-temporal features (\(x, y, t\)), which are transformed into hidden features and a predicted warp.
This warp guides \( C_{field} \) to produce warped spatial features, which are then concatenated with hidden features for the MLP in \( C_{field} \), creating RGB frames.
An enhancement module is applied to finalize the restoration, enhancing visual quality.

\subsection{3D Spatial-Temporal Deformation Field}
We construct a spatial-temporal feature space \( V \in \mathbb{R}^{Q \times M \times N \times T} \), storing Q-channel feature vectors \( V_{x_n,y_n,t_n} = \{V_q\}_{q=1}^{Q} \) at each location. The spatial dimensions \(x_n\) and \(y_n\) correspond to the video frame's height and width. A compact MLP transforms these feature vectors into predicted warp and hidden spatial-temporal features at \( (x_n, y_n, t_n) \).

Adopting a low-rank-decomposed representation similar to TensoRF \cite{Chen2022ECCV}, \( V \) is modeled using a 1D vector \( u \) for temporal variation and a full-rank matrix \( M \) for spatial variations in \( x \) and \( y \). The Q-channel feature vectors in \( u \) and \( M \) are dynamically updated during optimization.

To extract \( V_{x_n,y_n,t_n} \) at \( t_n \), we project \( (x_n, y_n) \) onto \( M \) and \( t_n \) onto \( u \), resulting in \( M_{x_n,y_n} \) and \( u_{t_n} \). The final feature vector is the Hadamard product:
\begin{equation}
    V_{x_n,y_n,t_n} = M_{x_n,y_n} \odot u_{t_n},
\end{equation}
where \( \odot \) is the Hadamard product. This efficiently approximates a 3D feature as a tensor product of a 2D matrix and 1D vector, reducing parameters while capturing spatial-temporal details. In the subsequent MLP, Layer normalization is enable to promote stable training.

\subsection{2D Content Field}
Within the 2D Content Field \( C_{field} \), we obtain spatial feature vectors \( M_{x_n,y_n} \) with the spatial coordinates \( (x_n, y_n) \). 
These vectors used to warped the grid points used to sample from \( D_{field} \), resulting in warped spatial features.
Subsequently, each warped vector is concatenated with hidden spatial-temporal features derived from \( D_{field} \). 
These concatenated features are then fed into the MLP and the subsequent enhancement module to produce the final RGB frame.

\subsection{Semantic Enhancement}
In our approach, the output from TurbNet \cite{mao2022single} serves as a solitary supervisory signal, which may inadvertently introduce artifacts into our results if relied upon exclusively. To enhance the quality of the visual output and imbue it with semantic depth, we utilize text-driven models, specifically CLIP, known for their proficiency in aligning visual content with textual descriptions. The Semantic Enhancement step is crucial for transcending mere fidelity, aiming instead for semantically enriched and contextually nuanced outputs.

Nevertheless, the arbitrary selection of text prompts could yield suboptimal results. To address this, we present a principled prompt-selection methodology, which employs statistical correlation between the Learned Perceptual Image Patch Similarity (LPIPS) and CLIP scores to select the most appropriate prompts. As demonstrated in Figure~\ref{fig:clip_select}, we conduct a comparative analysis of various degraded images using the same sequence of frames. We observe the relationship between LPIPS scores and CLIP losses, alongside a correlation study illustrated by the Kendall Rank Correlation Coefficient (KRCC) and Spearman's Rank Correlation Coefficient (SRCC) in the right figure of Figure~\ref{fig:clip_select}.

The selection process yields ``a degraded image with noise and turbulence distortion"  as the negative text prompt and ``a clean and sharp natural image" as the positive text prompt, which have shown the highest relevance in our tests. These prompts, meticulously chosen, are then integrated into our training as a new loss term:

\begin{equation}
\begin{aligned}
    \mathcal{L}_{text} = & -\left( \frac{\langle Enc_{i}(I), Enc_{t}(T_{pos}) \rangle}
    {\| Enc_{i}(I) \| \| Enc_{t}(T_{pos}) \|} \right. \\
    & \hspace{2em} \left. - \frac{\langle Enc_{i}(I), Enc_{t}(T_{neg}) \rangle}
    {\| Enc_{i}(I) \| \| Enc_{t}(T_{neg}) \|} \right).
\end{aligned}
\end{equation}
\( Enc_{i}(I) \) is the feature vector extracted from the predicted image \( I \) using CLIP's image encoder. \( Enc_{t}(T_{pos}) \) is the feature vector obtained from the positive prompt \( T_{pos} \) using CLIP's text encoder, and \( Enc_{t}(T_{neg}) \) is derived similarly from the negative prompt \( T_{neg} \).  With this term, we guide the learning process toward the direction of positive prompt semantics and away from the negative prompt semantics.

\subsection{Training Objectives}
\paragraph{Temporal Consistency Regularization.} To ensure temporal stability across video frames, 
we employ a disparity estimation network (MiDas \cite{ranftl2020towards}) and calculate pixel-wise disparities as weight for the predicted warp (one of \( D_{field} \)'s output) to maintain spatial consistency over time. The loss is defined as:
\begin{equation}
    \mathcal{L}_{temp} = (1 - \text{Disparity}(I)) \cdot \| \text{Predicted Warp} \|_1
\end{equation}
where \( \text{Disparity}(I) \) measures the pixel-level disparity, and \( \| \text{Predicted Warp} \|_1 \) enforces sparsity in the grid changes. The design of \( \mathcal{L}_{temp} \) minimizes the L1 norm of the predicted warp, conditioned by \( 1 - \text{Disparity}(I) \), to prioritize consistency in far regions based on the depth information. This focused approach on temporal consistency significantly reduces the propagation of turbulence-induced distortions, ensuring a smooth transition between frames.
\begin{figure}[t]
  \centering
  \includegraphics[width=\linewidth]{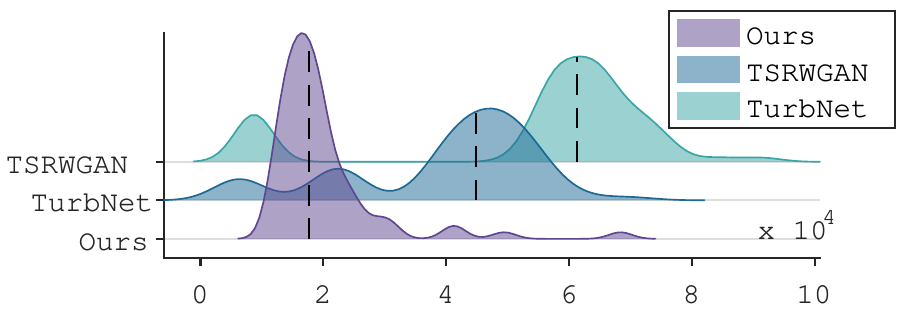}
    \caption{Histogram of TV for optical flow between frames in a video sequence with real-world turbulence. ConVRT obtains more defined optical flow, leading to lower total variation.}
   \label{fig:real_tv_score}
\end{figure}

\paragraph{Similarity Loss.} The Similarity Loss Term is given by:
\begin{equation}
\mathcal{L}_{sim} = \lambda_{mse} \mathcal{L}_{mse} + \lambda_{ssim} \mathcal{L}_{ssim} + \lambda_{lpips} \mathcal{L}_{lpips}
\end{equation}
where \( \lambda_{mse} \), \( \lambda_{ssim} \), and \( \lambda_{lpips} \) are weights for each term.
This loss term assesses the fidelity of the predicted output compared to TurbNet output, incorporating Mean Squared Error (MSE), Structural Similarity Index Measure (SSIM) \cite{wang2004image}, and Learned Perceptual Image Patch Similarity (LPIPS). This multifaceted approach ensures a comprehensive evaluation of reconstruction quality. 

\paragraph{Overall Loss}
The overall loss combines the similarity loss with temporal consistency and semantic enhancement:
\vspace{0.5em}
\begin{equation}
\mathcal{L}_{total} = \mathcal{L}_{sim} + \lambda_{temp} \mathcal{L}_{temp} + \lambda_{text} \mathcal{L}_{text}.
\end{equation}

\begin{table}[t]
\centering
\vspace{-3pt}
\caption{Evaluation metrics tested on video with synthetic turbulence. \(\uparrow\): higher is better, \(\downarrow\): lower is better.}

\label{table:runtime}
\vspace{-2pt}
\resizebox{\linewidth}{!}{%
\begin{tabular}{@{}cccccc@{}}
\toprule
Method           & \textbf{PSNR$_{Img}$ $\uparrow$} & \textbf{SSIM} $\uparrow$  & \textbf{LPIPS} $\downarrow$  & \textbf{E$_{warp}$} $\downarrow$ & \textbf{PSNR$_{x-t}$} $\uparrow$ \\ \hline 

TSRWGAN \cite{jin2021neutralizing} & 23.58 & 0.739 & 0.230  & 0.0026   & 23.77   \\ 
TurbNet \cite{mao2022single}       & 23.44 & 0.732 & 0.228  & 0.0057   & 23.54   \\ 
TurbNet+Real-ESRGAN \cite{wang2021real}  & 22.48 & 0.713 & 0.213  & 0.0074   & 22.67   \\
ConVRT (Ours)                               & \textbf{24.90} & \textbf{0.787} & \textbf{0.189}  & \textbf{0.0014}  & \textbf{25.73}   \\
\bottomrule 
\label{table:syn}
\end{tabular}%
}
\end{table}

\section{Experiments}
In this section, we provide the experimental details and results which validate the performance improvement enabled by our method.
{Additional experimental results are provided in the supplementary file.}

\begin{figure*}
  \centering
  \includegraphics[width=.95\linewidth]{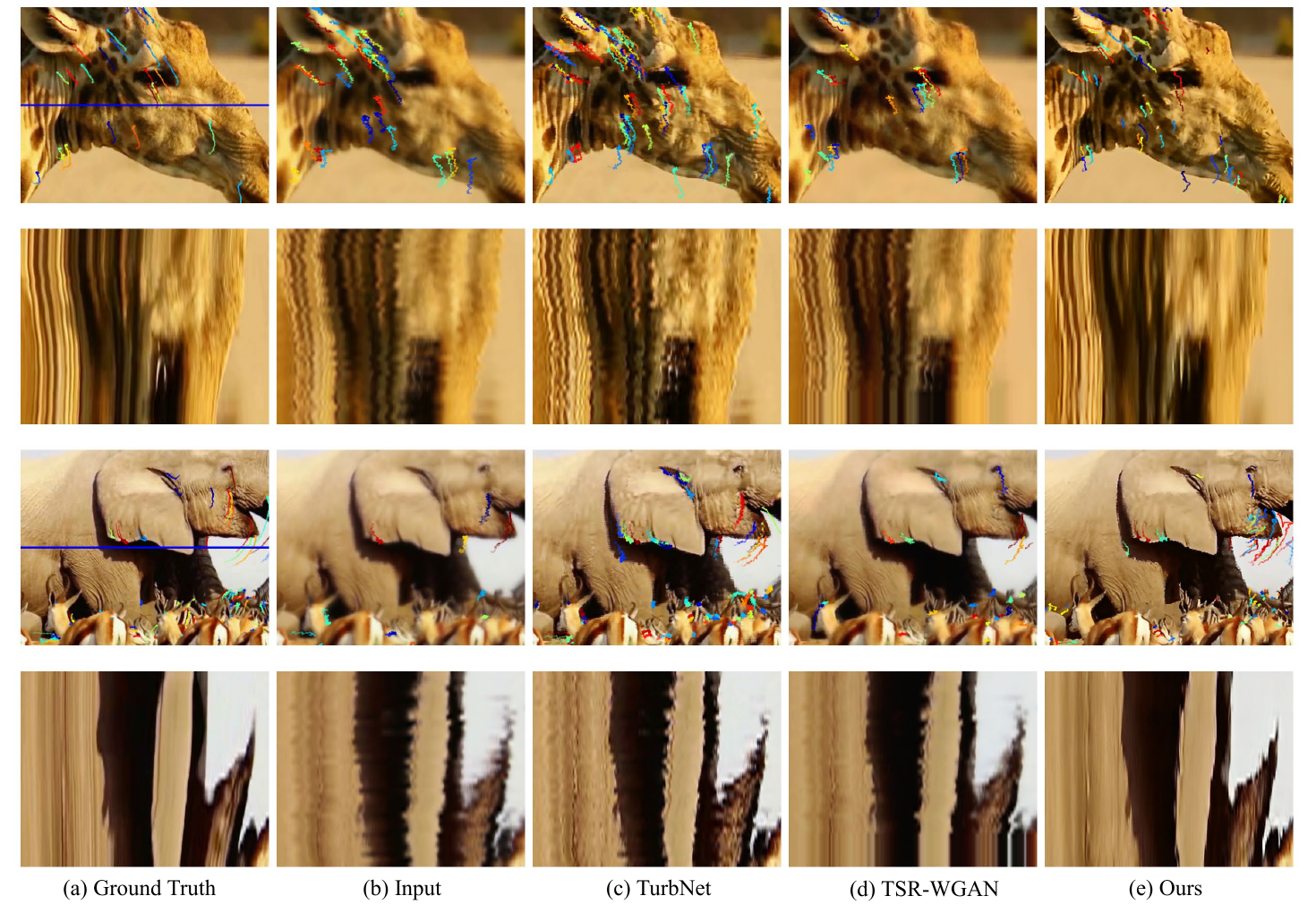}
    \caption{A comparison of synthetic turbulence video consistency: Two scenes are shown, each with two rows. The first row shows the KLT tracking trajectories, and the second shows an $x$-$t$ slice with a blue line indicating the slice's position. The ``zig-zag" patterns in TurbNet\cite{mao2022single} and TSR-WGAN\cite{jin2021neutralizing} trajectories show temporal inconsistency, and TSR-WGAN produces fewer trajectories, like the blurry input. In contrast, our method produces a smooth and reasonable number of trajectories. The $x$-$t$ slices from TurbNet and TSR-WGAN are non-smooth, whereas ours are smooth.}
    \label{fig:sim_temp_fig}
\end{figure*}

\begin{figure*}
  \centering
  \includegraphics[width=.96\linewidth]{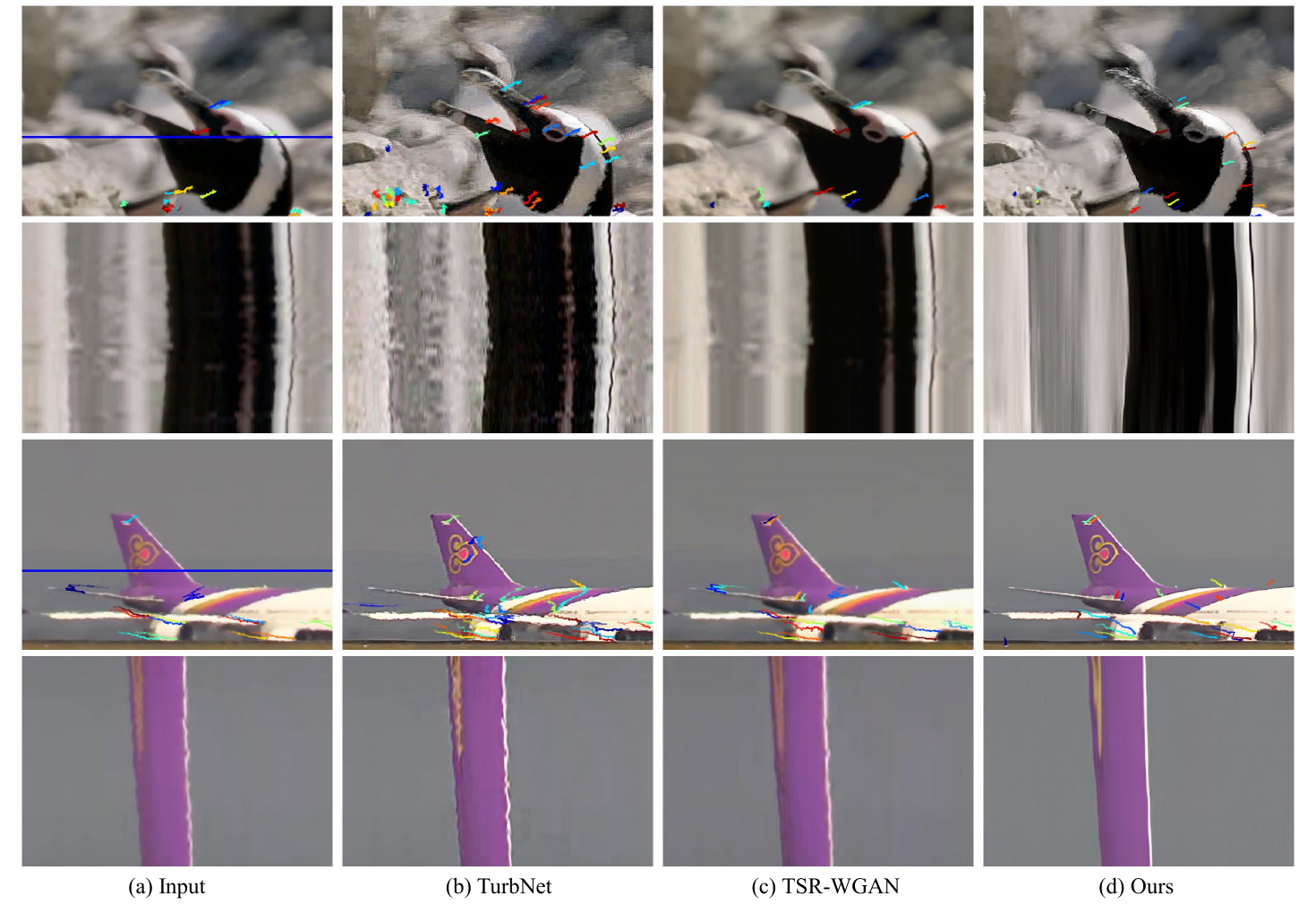}
   \caption{A comparison of temporal consistency in real-world turbulence videos with TurbNet\cite{mao2022single} and TSR-WGAN\cite{jin2021neutralizing}: Demonstrated in two scenes, each scene is represented by KLT tracking trajectories and an $x$-$t$ slice, indicated by a blue line. Our method shows the best temporal consistency in these real-world scenarios. Notably, in the first penguin scene with a stationary background, only our method accurately reflects stationary tracking trajectories (single-dot trajectories on the rock).}
    \label{fig:real_temp_fig}
\end{figure*}

\subsection{Datasets and Training Details}
We adopt subsets of the standard datasets, BVI-CLEAR dataset \cite{BVI-CLEAR} and the TSR-WGAN dataset \cite{jin2021neutralizing}, for fair comparison since baselines are trained among them. 
These datasets feature real-world turbulence and clean videos. 
To generate synthetic turbulence videos, we use the P2S atmospheric turbulence simulator \cite{mao2021accelerating} on clean real-world videos, producing our synthetic distorted video sequences. 
We employ turbulence parameters \( D/r_0 = 2 \) and \( corr = 1 \). 
Overall, this subset includes 8 sequences of real-world and synthetic videos with turbulence.
We train the ConVRT model for 6000 iterations with a learning rate of $2 \times 10^{-3}$.
The Adam optimizer \cite{kingma2014adam} is employed.
The enhancement module is Real-ESRGAN \cite{wang2021real}, which remains fixed during the training of ConVRT.

\subsection{Evaluation Strategy}
%
Two state-of-the-art methods of Turbulence Mitigation are used for fair comparison : TurbNet and TSRW-GAN. 
TMT \cite{zhang2022imaging}, the only video-based turbulence mitigation method, is skipped because the well-trained weight is inaccessible.
To evaluate the consistent removal of turbulence in video, we use four metrics for qualitative evaluation and two interframe-related methods for qualitative evaluation.

\paragraph{Per-frame Quality and Temporal Consistency.} We use PSNR and SSIM to measure the per-frame quality of reconstruction. 
LPIPS is used to measure the perceptual quality. Following \cite{jiabin}, we use the average warp error to measure the temporal consistency for the restored video. For the warp error between two consecutive frames, it can be defined as following:
\begin{align}
    E_{\text{warp}}(V_t, V_{t+1}) = \frac{1}{\sum_{i=1}^N M_t^{(i)}} \sum_{i=1}^N M_t^{(i)} \left\| V_t^{(i)} - \hat{V}_{t+1}^{(i)} \right\|^2_2,
\end{align}
where $\hat{V}_{t+1}^{(i)}$ is the warped frame by optical flow at time $t+1$ and $M_t^{(i)} \in \{0 , 1\}$ is the occlusion mask estimated by the methods proposed in \cite{ruder2016artistic}. The average warp error is:
\begin{align}
E_{\text{warp}}(V) = \frac{1}{T - 1} \sum_{t=1}^{T-1} E_{\text{warp}}(V_t, V_{t+1})
\end{align}
which is the average of consecutive warp errors across the entire video sequence.

\paragraph{KLT Trajectories.} We use the KLT tracker \cite{lucas1981iterative} to track the feature points, and then plot their trajectories as shown in Figure~\ref{fig:real_temp_fig}. 
KLT tracking is directly based on the image gradient information such that the common issues in turbulence restoration, i.e., blurriness, artifacts, temporal inconsistency, will be reflected in the tracked trajectories.
\paragraph{$x$-$t$ Slice.} We plot $x$-$t$ slices to show the motion of a row of pixels as shown in Figure \ref{fig:real_temp_fig}. 
If the video restoration is temporally inconsistent, the $x$-$t$ slice plot will show the non-smooth shape for curves.

\begin{table*}[t]
\centering
\vspace{-3pt}
\caption{Evaluation Text Prompts for Turbulence Mitigation Learning Guidance}
\label{table:runtime}
\vspace{-2pt}
\resizebox{\linewidth}{!}{%
\begin{tabular}{@{}cccccc@{}}
\toprule
 Text Index & Positive Prompts &   Negative Prompt \\ \hline
1 & ``a sharp image"   &   ``a blur image"    \\
2 & ``a sharp image"      & ``a image with blur and turbulence distortion"    \\
3 & ``a clean and sharp natural image"           & ``a degraded image with noise and turbulence distortion"    \\
4 & ``a clean and sharp natural image"    &  ``a degraded image with mosaic and turbulence distortion"    \\
5 & ``a clean and sharp natural image" &  ``a low-resolution image with mosaic and turbulence distortion" \\
6 & ``a clean and sharp natural image with table and alarm clock and books" & ``a low-resolution image with mosaic and turbulence distortion" \\
\bottomrule 

\label{table:clip}
\end{tabular}%
}
\end{table*}

\subsection{Comparison on Synthetic Turbulence Videos}
Our method achieves high temporal consistency while maintaining fidelity.
As observed in Figure \ref{fig:sim_single}, ConVRT surpasses other approaches in both turbulence removal and texture detail restoration. 
Figure \ref{fig:sim_temp_fig} demonstrates that the video dynamics generated by our method closely resemble those in the corresponding ground-truth videos (first column).
ConVRT uniquely captures the scene dynamics, effectively smoothing out atmospheric turbulence, a distinction not seen in other methods for turbulence mitigation.
The KLT trajectories further substantiate this temporal consistency. 
In contrast, TurbNet and TSR-WGAN produce ``zig-zag" tracking trajectories, indicative of temporally inconsistent video restoration. 
Notably, the KLT tracker generates only a few trajectories for TSR-WGAN's restoration and frame with turbulence (second column). 
This scarcity of trajectories might be attributed to this GAN-based method's generation of inconsistent content and failure to adequately deblur the scene across the video.
Table \ref{table:syn} presents these fidelity and temporal consistency metrics.
With the best per-frame quality, ConVRT outperforms all baseline methods in the temporal consistency metrics $E_{warp}$ and $PSNR$($x$-$t$).

\begin{figure}[t]
  \centering
  \includegraphics[width=\linewidth]{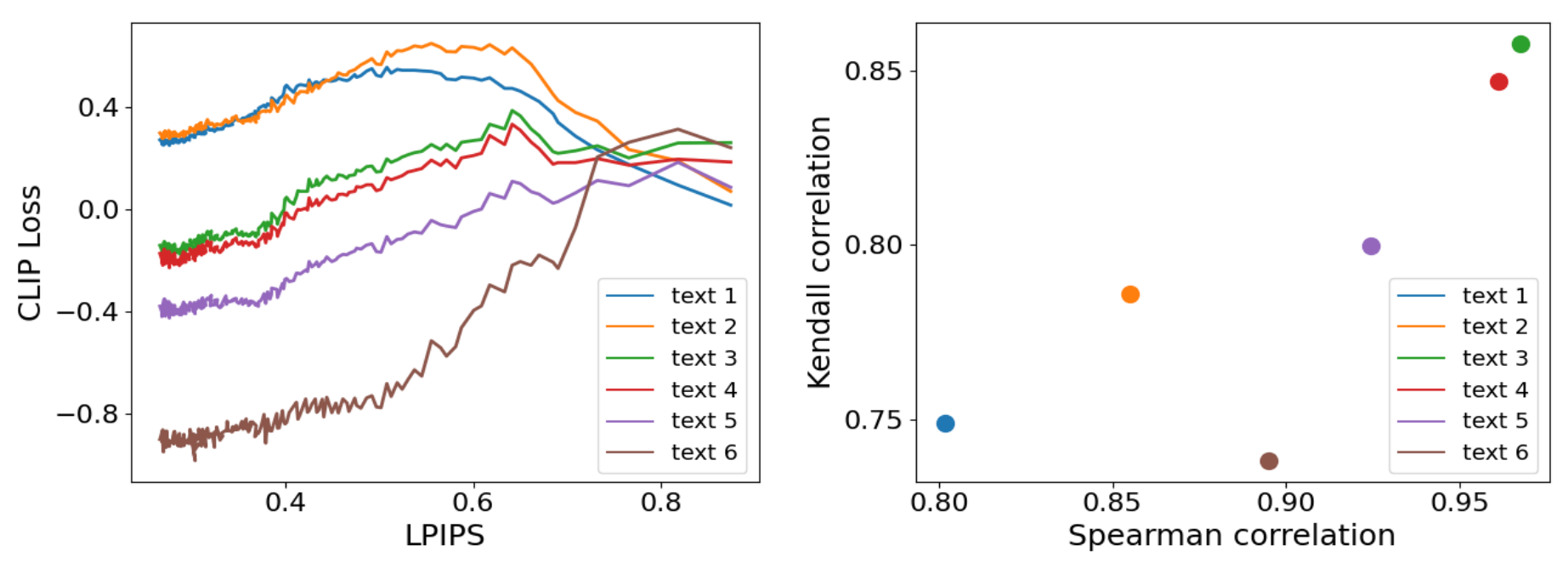}
   \caption{CLIP text prompt selection. Left: LPIPS against CLIP loss during optimization. Right: correlation scores between CLIP loss and LPIPS sequences for each text prompt. Text 3 is the final choice due to its highest correlation scores.}
   \label{fig:clip_select}
\end{figure}

\subsection{Comparison on Real-world Turbulence Videos}
Real-world atmospheric turbulence presents a significant and challenging domain gap compared to simulations.
However, our method achieves outstanding fidelity and temporal consistency in various video sequences distorted by real atmospheric turbulence, as demonstrated in the following figures.
Figure~\ref{fig:real_single} showcases the comparison of fidelity, where ConVRT restores outlines with sharper edge.

As in Figure~\ref{fig:real_temp_fig}, while the original turbulence and baseline methods result in ``zig-zag" tracking trajectories by KLT, ConVRT achieves smoother trajectories. This indicates that, unlike other methods, ConVRT consistently removes turbulence throughout the video. 
In addition, the x-t slice in Figure~\ref{fig:real_temp_fig} reveals that ConVRT smoothens row pixel motion more effectively, further enhancing single-frame turbulence removal (also shown in Figure \ref{fig:real_single}).

Figure~\ref{fig:real_tv_score} presents a histogram of Total Variation ($TV$) for optical flow between frames in a video with real-world turbulence. 
ConVRT attains lower TV, indicating more accurate optical flow, particularly on static backgrounds. 
Incorrect restoration of these backgrounds often leads to high TV due to chaotic optical flow.


\subsection{CLIP Text Prompt Selection}
Since we use the CLIP model to semantically guide the video restoration process, we investigate the effectiveness of different text prompts for guiding the turbulence removal and get better quality restoration. 
As shown in Figure \ref{fig:clip_select}, we use the ground-truth information to learn the turbulence removal process on a single image, we also simultaneously record the LPIPS learning curve and clip loss for different prompts from the Table \ref{table:clip}, then calculating the correlation between LPIPS sequence and clip loss sequences. The best text prompt is the text prompt that has loss with highest correlation score with LPIPS sequence. In this way we can quantitatively select the most correlated clip text prompt to guide the learning process of turbulence removal purpose.

\section{Discussion}
\paragraph{Difference to existing neural video representations?}
ConVRT does not merely represents the observed video, but actually restores the content before distorted by turbulence.
ConVRT accounts for the nuanced relationship between the temporal and spatial distortions of turbulence and adapting the neural representation to effectively model and counteract these distortions.
Conventional neural video representations do not account for such complex, dynamic distortions.

\paragraph{Difference between CLIP guidance to perceptual loss?}
Perceptual loss is less effective for turbulence mitigation as they typically require a reference clean image, which is often unavailable in turbulence-distorted scenarios. 
Our CLIP-guided module, on the other hand, bypasses this limitation by leveraging the pretrained CLIP model's ability to understand and interpret complex image content without needing a direct clean image reference, and yet allows for the integration of human priors through text prompts. 

\paragraph{Significance of improving the temporal consistency?}
Our method achieves smoother KLT tracking trajectories and maintains stationary backgrounds, which is crucial for downstream tasks like SLAM \cite{taketomi2017visual, aulinas2008slam, davison2007monoslam}, NeRF \cite{mildenhall2021nerf}, pose estimation \cite{schonberger2016structure}, and object segmentation and tracking \cite{yao2020video}. Improving temporal consistency in turbulence-mitigated videos will bring significant benefits to these tasks.
 
\paragraph{Why not use a simulator?}
Existing turbulence simulators may not accurately model distortions on backgrounds or distant objects. This is likely because 2D-based simulators (e.g., ~\cite{mao2021accelerating}) do not consider the depth effect on pixel displacement. 
Consequently, models based on these simulators struggle with far or background objects, as in Fig \ref{fig:real_temp_fig}, where baseline results show vibrating tracked points on stationary rocks. 
Simulating realistic turbulence for 3D scenes and objects are beyond the capacity of existing simulators. 

\section{Conclusion}

This paper presents ConVRT, a novel approach combining neural video representation with semantic supervision, enhanced by the pretrained CLIP model. ConVRT significantly improves temporal coherence and visual quality in video restoration, outperforming existing methods. ConVRT not only enhances video restoration quality under severe atmospheric turbulence, but also enables application scenarios like long-range object tracking and scene reconstruction. Overall, ConVRT represents a major step forward in long-range imaging, merging machine learning advancements to address key challenges and opening new avenues in computer vision and optical imaging.

\section*{Acknowledgements}
This work was supported in part by the AFOSR Young Investigator Program Award no. FA9550-22-1-0208 and ONR award no. N000142312752.
{
    \small
    \bibliographystyle{ieeenat_fullname}
    \bibliography{main}

\begin{thebibliography}{57}
\providecommand{\natexlab}[1]{#1}
\providecommand{\url}[1]{\texttt{#1}}
\expandafter\ifx\csname urlstyle\endcsname\relax
  \providecommand{\doi}[1]{doi: #1}\else
  \providecommand{\doi}{doi: \begingroup \urlstyle{rm}\Url}\fi

\bibitem[Anantrasirichai(2023)]{BVI-CLEAR}
Nantheera Anantrasirichai.
\newblock Atmospheric turbulence removal with complex-valued convolutional neural network.
\newblock \emph{Pattern Recognition Letters}, 171:\penalty0 69--75, 2023.

\bibitem[Anantrasirichai et~al.(2018)Anantrasirichai, Achim, and Bull]{anantrasirichai2018atmospheric}
Nantheera Anantrasirichai, Alin Achim, and David Bull.
\newblock Atmospheric turbulence mitigation for sequences with moving objects using recursive image fusion.
\newblock In \emph{2018 25th IEEE international conference on image processing (ICIP)}, pages 2895--2899. IEEE, 2018.

\bibitem[Attal et~al.(2023)Attal, Huang, Richardt, Zollhoefer, Kopf, O’Toole, and Kim]{attal2023hyperreel}
Benjamin Attal, Jia-Bin Huang, Christian Richardt, Michael Zollhoefer, Johannes Kopf, Matthew O’Toole, and Changil Kim.
\newblock Hyperreel: High-fidelity 6-dof video with ray-conditioned sampling.
\newblock In \emph{Proceedings of the IEEE/CVF Conference on Computer Vision and Pattern Recognition}, pages 16610--16620, 2023.

\bibitem[Aulinas et~al.(2008)Aulinas, Petillot, Salvi, and Llad{\'o}]{aulinas2008slam}
Josep Aulinas, Yvan Petillot, Joaquim Salvi, and Xavier Llad{\'o}.
\newblock The slam problem: a survey.
\newblock \emph{Artificial Intelligence Research and Development}, pages 363--371, 2008.

\bibitem[Bostan et~al.(2020)Bostan, Heckel, Chen, Kellman, and Waller]{bostan2020deep}
Emrah Bostan, Reinhard Heckel, Michael Chen, Michael Kellman, and Laura Waller.
\newblock Deep phase decoder: self-calibrating phase microscopy with an untrained deep neural network.
\newblock \emph{Optica}, 7\penalty0 (6):\penalty0 559--562, 2020.

\bibitem[Caliskan and Arica(2014)]{caliskan2014atmospheric}
Tufan Caliskan and Nafiz Arica.
\newblock Atmospheric turbulence mitigation using optical flow.
\newblock In \emph{2014 22nd International Conference on Pattern Recognition}, pages 883--888. Ieee, 2014.

\bibitem[Chen et~al.(2022)Chen, Xu, Geiger, Yu, and Su]{Chen2022ECCV}
Anpei Chen, Zexiang Xu, Andreas Geiger, Jingyi Yu, and Hao Su.
\newblock Tensorf: Tensorial radiance fields.
\newblock In \emph{European Conference on Computer Vision (ECCV)}, 2022.

\bibitem[Chen et~al.(2021{\natexlab{a}})Chen, He, Wang, Ren, Lim, and Shrivastava]{chen2021nerv}
Hao Chen, Bo He, Hanyu Wang, Yixuan Ren, Ser~Nam Lim, and Abhinav Shrivastava.
\newblock Nerv: Neural representations for videos.
\newblock \emph{Advances in Neural Information Processing Systems}, 34:\penalty0 21557--21568, 2021{\natexlab{a}}.

\bibitem[Chen et~al.(2023)Chen, Gwilliam, Lim, and Shrivastava]{chen2023hnerv}
Hao Chen, Matthew Gwilliam, Ser-Nam Lim, and Abhinav Shrivastava.
\newblock Hnerv: A hybrid neural representation for videos.
\newblock In \emph{Proceedings of the IEEE/CVF Conference on Computer Vision and Pattern Recognition}, pages 10270--10279, 2023.

\bibitem[Chen et~al.(2021{\natexlab{b}})Chen, Liu, and Wang]{chen2021learning}
Yinbo Chen, Sifei Liu, and Xiaolong Wang.
\newblock Learning continuous image representation with local implicit image function.
\newblock In \emph{Proceedings of the IEEE/CVF conference on computer vision and pattern recognition}, pages 8628--8638, 2021{\natexlab{b}}.

\bibitem[Chimitt and Chan(2023)]{chimitt2023anisoplanatic}
Nicholas Chimitt and Stanley~H Chan.
\newblock Anisoplanatic optical turbulence simulation for near-continuous profiles without wave propagation.
\newblock \emph{arXiv preprint arXiv:2305.09036}, 2023.

\bibitem[Chimitt et~al.(2023)Chimitt, Zhang, Chi, and Chan]{chimitt2023scattering}
Nicholas Chimitt, Xingguang Zhang, Yiheng Chi, and Stanley~H Chan.
\newblock Scattering and gathering for spatially varying blurs.
\newblock \emph{arXiv preprint arXiv:2303.05687}, 2023.

\bibitem[Davison et~al.(2007)Davison, Reid, Molton, and Stasse]{davison2007monoslam}
Andrew~J Davison, Ian~D Reid, Nicholas~D Molton, and Olivier Stasse.
\newblock Monoslam: Real-time single camera slam.
\newblock \emph{IEEE transactions on pattern analysis and machine intelligence}, 29\penalty0 (6):\penalty0 1052--1067, 2007.

\bibitem[Feng and Varshney(2021)]{feng2021signet}
Brandon~Yushan Feng and Amitabh Varshney.
\newblock Signet: Efficient neural representation for light fields.
\newblock In \emph{Proceedings of the IEEE/CVF International Conference on Computer Vision}, pages 14224--14233, 2021.

\bibitem[Feng and Varshney(2022)]{feng2022neural}
Brandon~Yushan Feng and Amitabh Varshney.
\newblock Neural subspaces for light fields.
\newblock \emph{IEEE Transactions on Visualization and Computer Graphics}, 2022.

\bibitem[Feng et~al.(2022{\natexlab{a}})Feng, Jabbireddy, and Varshney]{feng2022viinter}
Brandon~Yushan Feng, Susmija Jabbireddy, and Amitabh Varshney.
\newblock Viinter: View interpolation with implicit neural representations of images.
\newblock In \emph{SIGGRAPH Asia 2022 Conference Papers}, pages 1--9, 2022{\natexlab{a}}.

\bibitem[Feng et~al.(2022{\natexlab{b}})Feng, Xie, and Metzler]{feng2022turbugan}
Brandon~Y Feng, Mingyang Xie, and Christopher~A Metzler.
\newblock Turbugan: An adversarial learning approach to spatially-varying multiframe blind deconvolution with applications to imaging through turbulence.
\newblock \emph{IEEE Journal on Selected Areas in Information Theory}, 3\penalty0 (3):\penalty0 543--556, 2022{\natexlab{b}}.

\bibitem[Feng et~al.(2022{\natexlab{c}})Feng, Zhang, Tang, Du, and Varshney]{feng2022prif}
Brandon~Y Feng, Yinda Zhang, Danhang Tang, Ruofei Du, and Amitabh Varshney.
\newblock Prif: Primary ray-based implicit function.
\newblock In \emph{European Conference on Computer Vision}, pages 138--155. Springer, 2022{\natexlab{c}}.

\bibitem[Feng et~al.(2023{\natexlab{a}})Feng, Alzayer, Rubinstein, Freeman, and Huang]{feng20233d}
Brandon~Y Feng, Hadi Alzayer, Michael Rubinstein, William~T Freeman, and Jia-Bin Huang.
\newblock 3d motion magnification: Visualizing subtle motions from time-varying radiance fields.
\newblock In \emph{Proceedings of the IEEE/CVF International Conference on Computer Vision}, pages 9837--9846, 2023{\natexlab{a}}.

\bibitem[Feng et~al.(2023{\natexlab{b}})Feng, Guo, Xie, Boominathan, Sharma, Veeraraghavan, and Metzler]{feng2023neuws}
Brandon~Y Feng, Haiyun Guo, Mingyang Xie, Vivek Boominathan, Manoj~K Sharma, Ashok Veeraraghavan, and Christopher~A Metzler.
\newblock Neuws: Neural wavefront shaping for guidestar-free imaging through static and dynamic scattering media.
\newblock \emph{Science Advances}, 9\penalty0 (26):\penalty0 eadg4671, 2023{\natexlab{b}}.

\bibitem[Fried(1978)]{fried1978probability}
David~L Fried.
\newblock Probability of getting a lucky short-exposure image through turbulence.
\newblock \emph{JOSA}, 68\penalty0 (12):\penalty0 1651--1658, 1978.

\bibitem[Gilles et~al.(2008)Gilles, Dagobert, and De~Franchis]{gilles2008atmospheric}
J{\'e}r{\^o}me Gilles, Tristan Dagobert, and Carlo De~Franchis.
\newblock Atmospheric turbulence restoration by diffeomorphic image registration and blind deconvolution.
\newblock In \emph{Advanced Concepts for Intelligent Vision Systems: 10th International Conference, ACIVS 2008, Juan-les-Pins, France, October 20-24, 2008. Proceedings 10}, pages 400--409. Springer, 2008.

\bibitem[Jiang et~al.(2023)Jiang, Boominathan, and Veeraraghavan]{jiang2023nert}
Weiyun Jiang, Vivek Boominathan, and Ashok Veeraraghavan.
\newblock Nert: Implicit neural representations for unsupervised atmospheric turbulence mitigation.
\newblock In \emph{Proceedings of the IEEE/CVF Conference on Computer Vision and Pattern Recognition}, pages 4235--4242, 2023.

\bibitem[Jin et~al.(2021)Jin, Chen, Lu, Chen, Wang, Liu, Guo, and Bai]{jin2021neutralizing}
Darui Jin, Ying Chen, Yi Lu, Junzhang Chen, Peng Wang, Zichao Liu, Sheng Guo, and Xiangzhi Bai.
\newblock Neutralizing the impact of atmospheric turbulence on complex scene imaging via deep learning.
\newblock \emph{Nature Machine Intelligence}, 3\penalty0 (10):\penalty0 876--884, 2021.

\bibitem[Kasten et~al.(2021)Kasten, Ofri, Wang, and Dekel]{kasten2021layered}
Yoni Kasten, Dolev Ofri, Oliver Wang, and Tali Dekel.
\newblock Layered neural atlases for consistent video editing.
\newblock \emph{ACM Transactions on Graphics (TOG)}, 40\penalty0 (6):\penalty0 1--12, 2021.

\bibitem[Kingma and Ba(2014)]{kingma2014adam}
Diederik~P Kingma and Jimmy Ba.
\newblock Adam: A method for stochastic optimization.
\newblock \emph{arXiv preprint arXiv:1412.6980}, 2014.

\bibitem[Lai et~al.(2018)Lai, Huang, Wang, Shechtman, Yumer, and Yang]{jiabin}
Wei-Sheng Lai, Jia-Bin Huang, Oliver Wang, Eli Shechtman, Ersin Yumer, and Ming-Hsuan Yang.
\newblock Learning blind video temporal consistency.
\newblock In \emph{Proceedings of the European conference on computer vision (ECCV)}, pages 170--185, 2018.

\bibitem[Lee et~al.(2023)Lee, Jang, Chen, Qiu, and Huang]{lee2023shape}
Yao-Chih Lee, Ji-Ze~Genevieve Jang, Yi-Ting Chen, Elizabeth Qiu, and Jia-Bin Huang.
\newblock Shape-aware text-driven layered video editing.
\newblock In \emph{Proceedings of the IEEE/CVF Conference on Computer Vision and Pattern Recognition}, pages 14317--14326, 2023.

\bibitem[Li et~al.(2023)Li, Wang, Cole, Tucker, and Snavely]{li2023dynibar}
Zhengqi Li, Qianqian Wang, Forrester Cole, Richard Tucker, and Noah Snavely.
\newblock Dynibar: Neural dynamic image-based rendering.
\newblock In \emph{Proceedings of the IEEE/CVF Conference on Computer Vision and Pattern Recognition}, pages 4273--4284, 2023.

\bibitem[Lin et~al.(2023)Lin, Wang, Lin, Miau, Kainz, Chen, Zhang, Lindell, and Kutulakos]{lin2023learning}
Esther~YH Lin, Zhecheng Wang, Rebecca Lin, Daniel Miau, Florian Kainz, Jiawen Chen, Xuaner~Cecilia Zhang, David~B Lindell, and Kiriakos~N Kutulakos.
\newblock Learning lens blur fields.
\newblock \emph{arXiv preprint arXiv:2310.11535}, 2023.

\bibitem[Lucas and Kanade(1981)]{lucas1981iterative}
Bruce~D Lucas and Takeo Kanade.
\newblock An iterative image registration technique with an application to stereo vision.
\newblock In \emph{IJCAI'81: 7th international joint conference on Artificial intelligence}, pages 674--679, 1981.

\bibitem[Mao et~al.(2020)Mao, Chimitt, and Chan]{mao2020image}
Zhiyuan Mao, Nicholas Chimitt, and Stanley~H Chan.
\newblock Image reconstruction of static and dynamic scenes through anisoplanatic turbulence.
\newblock \emph{IEEE Transactions on Computational Imaging}, 6:\penalty0 1415--1428, 2020.

\bibitem[Mao et~al.(2021)Mao, Chimitt, and Chan]{mao2021accelerating}
Zhiyuan Mao, Nicholas Chimitt, and Stanley~H Chan.
\newblock Accelerating atmospheric turbulence simulation via learned phase-to-space transform.
\newblock In \emph{Proceedings of the IEEE/CVF International Conference on Computer Vision}, pages 14759--14768, 2021.

\bibitem[Mao et~al.(2022)Mao, Jaiswal, Wang, and Chan]{mao2022single}
Zhiyuan Mao, Ajay Jaiswal, Zhangyang Wang, and Stanley~H Chan.
\newblock Single frame atmospheric turbulence mitigation: A benchmark study and a new physics-inspired transformer model.
\newblock In \emph{European Conference on Computer Vision}, pages 430--446. Springer, 2022.

\bibitem[Mehta et~al.(2021)Mehta, Gharbi, Barnes, Shechtman, Ramamoorthi, and Chandraker]{mehta2021modulated}
Ishit Mehta, Micha{\"e}l Gharbi, Connelly Barnes, Eli Shechtman, Ravi Ramamoorthi, and Manmohan Chandraker.
\newblock Modulated periodic activations for generalizable local functional representations.
\newblock In \emph{Proceedings of the IEEE/CVF International Conference on Computer Vision}, pages 14214--14223, 2021.

\bibitem[Mildenhall et~al.(2021)Mildenhall, Srinivasan, Tancik, Barron, Ramamoorthi, and Ng]{mildenhall2021nerf}
Ben Mildenhall, Pratul~P Srinivasan, Matthew Tancik, Jonathan~T Barron, Ravi Ramamoorthi, and Ren Ng.
\newblock Nerf: Representing scenes as neural radiance fields for view synthesis.
\newblock \emph{Communications of the ACM}, 65\penalty0 (1):\penalty0 99--106, 2021.

\bibitem[M{\"u}ller et~al.(2022)M{\"u}ller, Evans, Schied, and Keller]{muller2022instant}
Thomas M{\"u}ller, Alex Evans, Christoph Schied, and Alexander Keller.
\newblock Instant neural graphics primitives with a multiresolution hash encoding.
\newblock \emph{ACM Transactions on Graphics (ToG)}, 41\penalty0 (4):\penalty0 1--15, 2022.

\bibitem[Noll(1976)]{noll1976zernike}
Robert~J Noll.
\newblock Zernike polynomials and atmospheric turbulence.
\newblock \emph{JOsA}, 66\penalty0 (3):\penalty0 207--211, 1976.

\bibitem[Ouyang et~al.(2023)Ouyang, Wang, Xiao, Bai, Zhang, Zheng, Zhou, Chen, and Shen]{ouyang2023codef}
Hao Ouyang, Qiuyu Wang, Yuxi Xiao, Qingyan Bai, Juntao Zhang, Kecheng Zheng, Xiaowei Zhou, Qifeng Chen, and Yujun Shen.
\newblock Codef: Content deformation fields for temporally consistent video processing.
\newblock \emph{arXiv preprint arXiv:2308.07926}, 2023.

\bibitem[Park et~al.(2019)Park, Florence, Straub, Newcombe, and Lovegrove]{park2019deepsdf}
Jeong~Joon Park, Peter Florence, Julian Straub, Richard Newcombe, and Steven Lovegrove.
\newblock Deepsdf: Learning continuous signed distance functions for shape representation.
\newblock In \emph{Proceedings of the IEEE/CVF conference on computer vision and pattern recognition}, pages 165--174, 2019.

\bibitem[Radford et~al.(2021)Radford, Kim, Hallacy, Ramesh, Goh, Agarwal, Sastry, Askell, Mishkin, Clark, et~al.]{radford2021learning}
Alec Radford, Jong~Wook Kim, Chris Hallacy, Aditya Ramesh, Gabriel Goh, Sandhini Agarwal, Girish Sastry, Amanda Askell, Pamela Mishkin, Jack Clark, et~al.
\newblock Learning transferable visual models from natural language supervision.
\newblock In \emph{International conference on machine learning}, pages 8748--8763. PMLR, 2021.

\bibitem[Ranftl et~al.(2020)Ranftl, Lasinger, Hafner, Schindler, and Koltun]{ranftl2020towards}
Ren{\'e} Ranftl, Katrin Lasinger, David Hafner, Konrad Schindler, and Vladlen Koltun.
\newblock Towards robust monocular depth estimation: Mixing datasets for zero-shot cross-dataset transfer.
\newblock \emph{IEEE transactions on pattern analysis and machine intelligence}, 44\penalty0 (3):\penalty0 1623--1637, 2020.

\bibitem[Ruder et~al.(2016)Ruder, Dosovitskiy, and Brox]{ruder2016artistic}
Manuel Ruder, Alexey Dosovitskiy, and Thomas Brox.
\newblock Artistic style transfer for videos.
\newblock In \emph{Pattern Recognition: 38th German Conference, GCPR 2016, Hannover, Germany, September 12-15, 2016, Proceedings 38}, pages 26--36. Springer, 2016.

\bibitem[Schonberger and Frahm(2016)]{schonberger2016structure}
Johannes~L Schonberger and Jan-Michael Frahm.
\newblock Structure-from-motion revisited.
\newblock In \emph{Proceedings of the IEEE conference on computer vision and pattern recognition}, pages 4104--4113, 2016.

\bibitem[Shimizu et~al.(2008)Shimizu, Yoshimura, Tanaka, and Okutomi]{shimizu2008super}
Masao Shimizu, Shin Yoshimura, Masayuki Tanaka, and Masatoshi Okutomi.
\newblock Super-resolution from image sequence under influence of hot-air optical turbulence.
\newblock In \emph{2008 IEEE Conference on Computer Vision and Pattern Recognition}, pages 1--8. IEEE, 2008.

\bibitem[Sitzmann et~al.(2020)Sitzmann, Martel, Bergman, Lindell, and Wetzstein]{sitzmann2020implicit}
Vincent Sitzmann, Julien Martel, Alexander Bergman, David Lindell, and Gordon Wetzstein.
\newblock Implicit neural representations with periodic activation functions.
\newblock \emph{Advances in neural information processing systems}, 33:\penalty0 7462--7473, 2020.

\bibitem[Taketomi et~al.(2017)Taketomi, Uchiyama, and Ikeda]{taketomi2017visual}
Takafumi Taketomi, Hideaki Uchiyama, and Sei Ikeda.
\newblock Visual slam algorithms: A survey from 2010 to 2016.
\newblock \emph{IPSJ Transactions on Computer Vision and Applications}, 9\penalty0 (1):\penalty0 1--11, 2017.

\bibitem[Tancik et~al.(2020)Tancik, Srinivasan, Mildenhall, Fridovich-Keil, Raghavan, Singhal, Ramamoorthi, Barron, and Ng]{tancik2020fourier}
Matthew Tancik, Pratul Srinivasan, Ben Mildenhall, Sara Fridovich-Keil, Nithin Raghavan, Utkarsh Singhal, Ravi Ramamoorthi, Jonathan Barron, and Ren Ng.
\newblock Fourier features let networks learn high frequency functions in low dimensional domains.
\newblock \emph{Advances in Neural Information Processing Systems}, 33:\penalty0 7537--7547, 2020.

\bibitem[Wang and Tian(2023)]{wang2023local}
Hao Wang and Lei Tian.
\newblock Local conditional neural fields for versatile and generalizable large-scale reconstructions in computational imaging.
\newblock \emph{arXiv preprint arXiv:2307.06207}, 2023.

\bibitem[Wang et~al.(2023)Wang, Chang, Cai, Li, Hariharan, Holynski, and Snavely]{wang2023tracking}
Qianqian Wang, Yen-Yu Chang, Ruojin Cai, Zhengqi Li, Bharath Hariharan, Aleksander Holynski, and Noah Snavely.
\newblock Tracking everything everywhere all at once.
\newblock \emph{arXiv preprint arXiv:2306.05422}, 2023.

\bibitem[Wang et~al.(2021)Wang, Xie, Dong, and Shan]{wang2021real}
Xintao Wang, Liangbin Xie, Chao Dong, and Ying Shan.
\newblock Real-esrgan: Training real-world blind super-resolution with pure synthetic data.
\newblock In \emph{Proceedings of the IEEE/CVF international conference on computer vision}, pages 1905--1914, 2021.

\bibitem[Wang et~al.(2004)Wang, Bovik, Sheikh, and Simoncelli]{wang2004image}
Zhou Wang, Alan~C Bovik, Hamid~R Sheikh, and Eero~P Simoncelli.
\newblock Image quality assessment: from error visibility to structural similarity.
\newblock \emph{IEEE transactions on image processing}, 13\penalty0 (4):\penalty0 600--612, 2004.

\bibitem[Yao et~al.(2020)Yao, Lin, Xia, Zhao, and Zhou]{yao2020video}
Rui Yao, Guosheng Lin, Shixiong Xia, Jiaqi Zhao, and Yong Zhou.
\newblock Video object segmentation and tracking: A survey.
\newblock \emph{ACM Transactions on Intelligent Systems and Technology (TIST)}, 11\penalty0 (4):\penalty0 1--47, 2020.

\bibitem[Ye et~al.(2022)Ye, Li, Tucker, Kanazawa, and Snavely]{ye2022deformable}
Vickie Ye, Zhengqi Li, Richard Tucker, Angjoo Kanazawa, and Noah Snavely.
\newblock Deformable sprites for unsupervised video decomposition.
\newblock In \emph{Proceedings of the IEEE/CVF Conference on Computer Vision and Pattern Recognition}, pages 2657--2666, 2022.

\bibitem[Zhang et~al.(2018)Zhang, Isola, Efros, Shechtman, and Wang]{zhang2018unreasonable}
Richard Zhang, Phillip Isola, Alexei~A Efros, Eli Shechtman, and Oliver Wang.
\newblock The unreasonable effectiveness of deep features as a perceptual metric.
\newblock In \emph{Proceedings of the IEEE conference on computer vision and pattern recognition}, pages 586--595, 2018.

\bibitem[Zhang et~al.(2022)Zhang, Mao, Chimitt, and Chan]{zhang2022imaging}
Xingguang Zhang, Zhiyuan Mao, Nicholas Chimitt, and Stanley~H Chan.
\newblock Imaging through the atmosphere using turbulence mitigation transformer.
\newblock \emph{arXiv preprint arXiv:2207.06465}, 2022.

\bibitem[Zhou et~al.(2023)Zhou, Feng, Guo, Liang, Metzler, Yang, et~al.]{zhou2023fpm}
Haowen Zhou, Brandon~Y Feng, Haiyun Guo, Mingshu Liang, Christopher~A Metzler, Changhuei Yang, et~al.
\newblock Fpm-inr: Fourier ptychographic microscopy image stack reconstruction using implicit neural representations.
\newblock \emph{arXiv preprint arXiv:2310.18529}, 2023.

\end{thebibliography}
}
\end{document}


\twocolumn[{
\maketitle
\begin{center}
    \captionsetup{type=figure}
    \includegraphics[width=\linewidth\baselineskip]{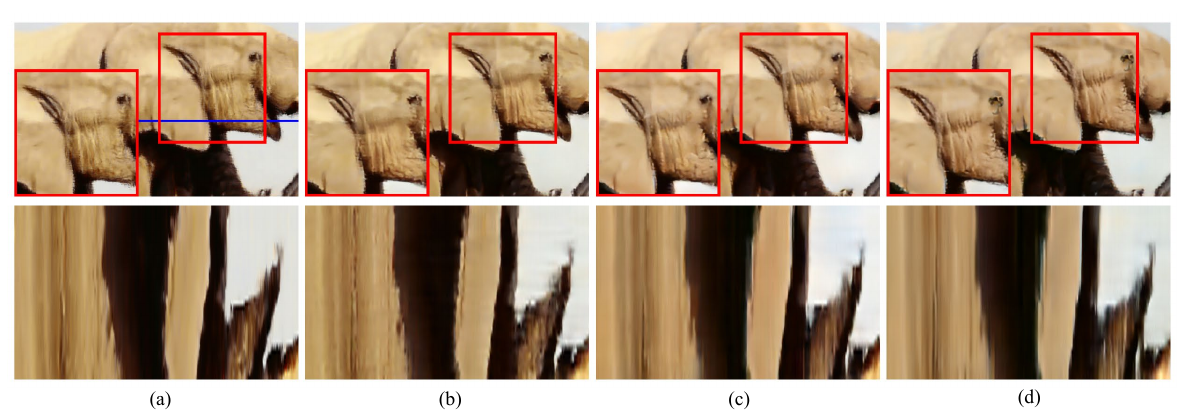}
    \captionof{figure}{Ablation study showcases effect of proposed components. The top row demonstrates sharpness across the proposed components, while the bottom row compares temporal consistency through $x$-$t$ slices. Column (a) presents the base of ConVRT model (comprising \( C_{field} \) and \( D_{field} \)). Column (b) depicts the enhancement in sharpness due to the addition of Real-ESRGAN. Column (c) shows the further refinement of image detail with the integration of CLIP. Lastly, column (d) illustrates the improved temporal consistency achieved by implementing the temporal consistency loss. The performance corresponding to each component, as evaluated by various metrics, is presented in Table \ref{table:abala}.}
    \label{fig:ablation}
\end{center}
}]

\maketitle
\thispagestyle{empty}
\appendix

\begin{figure*}[t]
  \centering
  \includegraphics[width=\linewidth]{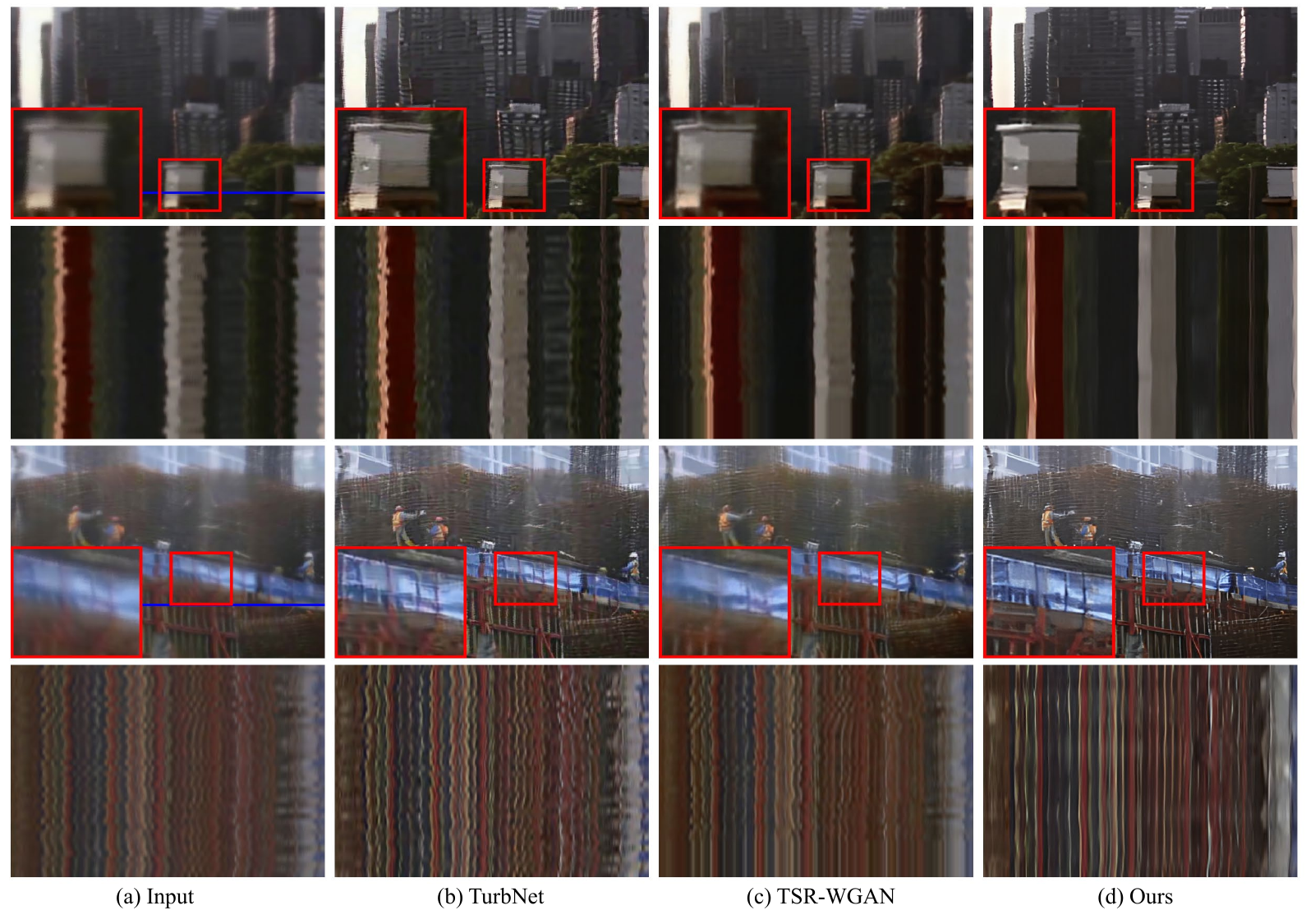}
   \caption{More comparisons on synthetic turbulence video. Two scenes are presented, with sharpness comparisons in the first row and temporal consistency comparisons in the second row through $x$-$t$ slices. ConVRT exhibits smooth $x$-$t$ slice patterns, indicating superior temporal stability over TurbNet and TSR-WGAN. ConVRT consistently produces sharper edges across the overall and zoomed-in regions.}
   \label{fig:supp_syn_more_1}
\end{figure*}
\begin{figure*}[t]
  \centering
  \includegraphics[width=\linewidth]{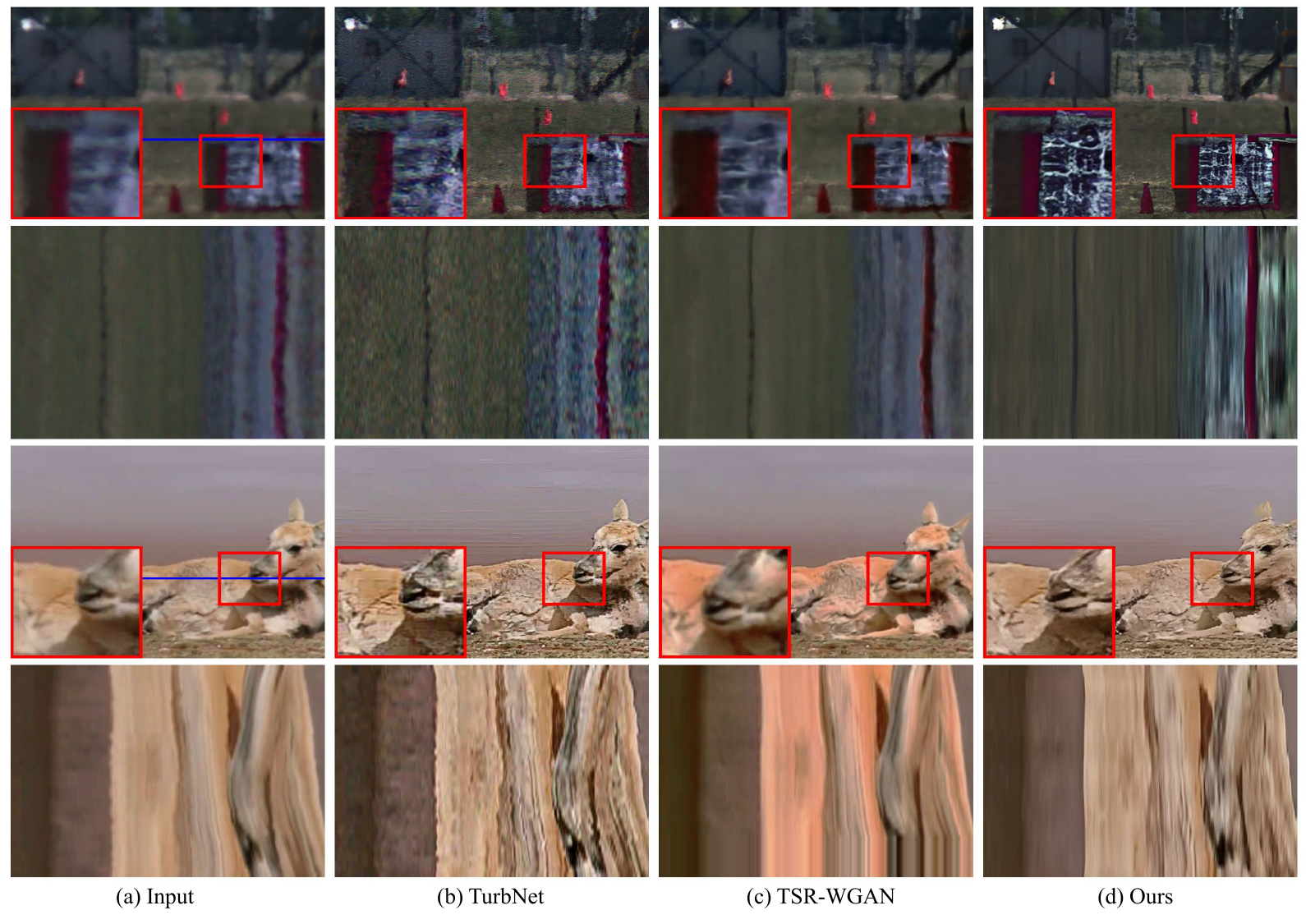}
   \caption{More comparisons on real-world turbulence video. Two scenes are presented, with sharpness comparisons in the first row and temporal consistency comparisons in the second row through $x$-$t$ slices. }
   \label{fig:supp_real_more_1}
\end{figure*}


\section{More Qualitative Comparison}
\textbf{The supplementary material includes an  \textcolor{red}{HTML page containing video results}.}
Please check the offline page for additional video results.
Additional visual comparisons on synthetic and real-world turbulence videos are provided in Fig \ref{fig:supp_syn_more_1} and Fig \ref{fig:supp_real_more_1}, respectively.

\begin{table}[t]
\centering
\vspace{-3pt}
\caption{Ablation Study : The table presents PSNR$_{\text{img}}$, SSIM, LPIPS, E$_{\text{warp}}$, and PSNR$_{x-t}$ scores, demonstrating the improvements achieved by each successive refinement, culminating in our proposed ConVRT method. (a), (b), (c), and (d) correspond to the respective indices in Fig \textcolor{red}{9}.}
\vspace{-2pt}
\resizebox{\linewidth}{!}{%
\begin{tabular}{@{}lccccc@{}}
\toprule
Method                          & \textbf{PSNR$_{Img}$ $\uparrow$} & \textbf{SSIM} $\uparrow$  & \textbf{LPIPS} $\downarrow$  & \textbf{E$_{warp}$} $\downarrow$ & \textbf{PSNR$_{x-t}$} $\uparrow$ \\ \hline
TSRWGAN \cite{jin2021neutralizing} & 23.58 & 0.739 & 0.230  & 0.0026   & 23.77   \\ 
TurbNet \cite{mao2022single}       & 23.44 & 0.732 & 0.228  & 0.0057   & 23.54   \\ 
TurbNet+Real-ESRGAN \cite{wang2021real}  & 22.48 & 0.713 & 0.213  & 0.0074   & 22.67   \\ \hline
Base (a)                             & 24.91 & 0.781 & 0.209  & 0.0017  & 25.44    \\
\quad + Real-ESRGAN (b)                     & 24.88 & 0.782 & 0.204  & 0.0020  & 25.11    \\
\quad + CLIP (c)                     & 24.71 & 0.784 & 0.193  & 0.0014  & 25.40    \\
\quad + temp. consistency (d)          & 24.90 & 0.787 & 0.189  & 0.0014  & 25.73    \\
= Ours (ConVRT) (d)                  & \textbf{24.90} & \textbf{0.787} & \textbf{0.189}  & \textbf{0.0014}  & \textbf{25.73}   \\
\bottomrule 

\label{table:abala}

\end{tabular}%
}
\end{table}
\vspace{-8pt}



\section{Ablation Study}

In this section, we assess the impact of each component added to our baseline model, as shown in the Fig \ref{fig:ablation} and quantified in Table \ref{table:abala}. 
Our base (the column \( (a) \) of Fig \ref{fig:ablation}), consisting solely of the pure framework (represented by \( C_{field} \) and \( D_{field} \)), serves as the starting point for this ablation study.
Even without the cooperative components, our baseline method in ablation study already exhibits notable performance in sharpness and temporal consistency, surpassing that of TurbuNet and TurbuNet enhanced with Real-ESRGAN.
The incorporation of Real-ESRGAN, visible in the column \( (b) \) of Fig \ref{fig:ablation}, improves the LPIPS score, indicating better perceptual image quality. 
This is evident from the improved sharpness and texture definition.
It's important to note that the naive combination of Real-ESRGAN and TurbuNet by themselves does not attain the level of temporal consistency that our approach does. 
This is reflected by the \( \mathbf{E}_{\text{warp}} \) metric detailed in Table \ref{table:abala}.
Adding CLIP, as shown in the column \( (c) \) of Fig \ref{fig:ablation}, further improves the LPIPS score, suggesting that the images have become sharper and more detailed. 
Finally, adding the temporal consistency regularization, as depicted in the column \( (d) \) of Fig \ref{fig:ablation}, not only enhances the consistency metric but also visibly improves the stability of the images over time. 
These steps clearly demonstrate the effectiveness of each proposed component in improving the sharpness and temporal consistency in our restoration.

{
    \small
    \bibliographystyle{ieeenat_fullname}
    \bibliography{main}
}